\documentclass[aps,showpacs,epsfig,twocolumn]{revtex4}
\usepackage{epsfig}

\newcommand{\be}{\begin{equation}}
\newcommand{\ee}{\end{equation}}
\newcommand{\bea}{\begin{eqnarray}}
\newcommand{\eea}{\end{eqnarray}}
\usepackage{color}
\begin{document}

\title{\bf\Large {Ferromagnetism of $UGe_2$}}

\author{Naoum Karchev }

\affiliation{Department of Physics, University of Sofia, 1126 Sofia, Bulgaria }

\begin{abstract}

Magnetism of $UGe_2$ is due to the magnetic ordered moments of $5f$ uranium electrons.
The strong spin-orbit coupling splits them into two groups.
The magnetization is investigated in terms of two vector fields
${\bf M}_{1i}$ and ${\bf M}_{2i}$ which identify the local orientation of the magnetization
of the two groups of $f$ electrons.
Renormalized spin-wave theory, which accounts for the magnon-magnon interaction,
and its extension are developed to describe two ferromagnetic phases in
the system: low temperature large moment
phase  $0<T<T^{*}$ (FM2), where all $5f$ electrons contribute the ordered ferromagnetic
moment, and high temperature low-moment phase $T^{*}<T<T_C$ (FM1), where $f$ electrons are
partially ordered. Both phases are strictly ferromagnetic in accordance with experiment.
The  magnetization as a function of temperature is calculated. The anomalous temperature dependence
of the ordered moment, known from the experiments with $UGe_2$, is very well reproduced theoretically.
Below $T_x$ ($T^*$ in the present paper) the ferromagnetic moment increases in an anomalous way. The new
understanding of the anomalous $FM2\rightarrow FM1$ transition, as a result of the magnetic order of two well
separated groups of $f$ electrons, yields the key to an understanding of
the ferromagnetism and transport properties in these compounds.

\end{abstract}

\maketitle

The uranium compound $UGe_2$ is a metallic ferromagnet below the Curie temperature $T_C=52 K$ at
ambient pressure with a zero temperature ordered moment $\mu_s=1.48\mu_B/U$ \cite{Onuki92}.
The experimental measurements reveal the presence of an
additional phase line that lies entirely within the ferromagnetic
phase. The characteristic temperature of this transition $T_{x}$,
which is below the Curie temperature $T_C$, decreases with pressure
and disappears at a pressure close to the pressure at which new
phase of coexistence of superconductivity and ferromagnetism
emerges\cite{2fmp1,2fmp2,2fmp3}. The additional phase transition
demonstrates itself through the change in the $T$ dependence of
the ordered ferromagnetic moment\cite{2fmp5,2fmp6,2fmp7}. The
magnetization shows an anomalous enhancement below $T_{x}$.

Magnetism of $UGe_2$ is due to the magnetic ordered moments of $5f$ uranium electrons.
They have dual character and in $UGe_2$ are more
itinerant than in many uranium compounds known as "heavy-fermion systems".
The degree of delocalization of the $5f$ electrons has been explored by variety of experimental
techniques. The thermodynamic properties, such as the magnetoresistance,
suggest that uranium $5f$ electrons behave like $3d$
electrons in the conventional itinerant ferromagnets \cite{Onuki92}. On the other hand, the
inelastic scattering experiments suggest localized character of the $5f$ electrons in
 $UGe_2$ \cite{2fmp2,2fmp3}. Finally the itinerant ferromagnetism may be inferred from the fact,
 that $UGe_2$ forms a very good metal. High quality single crystals have residual resistivity
 well below $1\mu\Omega cm$\cite{Pfleiderer2}.

 Because of the strong spin-orbit coupling of $f$ electrons one has to label the states
 by the total angular momentum $\textbf{J}=\textbf{L}+\textbf{S}$, where $\textbf{L}$ and
 $\textbf{S}$ are angular and spin momenta respectively. For $f$-orbitals $l=3$ and one obtains
 two multiplets, an octet with $j=7/2$ and a sextet with $j=5/2$. They are well separated by
 the spin-orbit interaction and the energy level of the octet is higher than the sextet one. Since
 the number of $f$ electrons in $UGe_2$ is less than six,  it is enough to consider $j=5/2$ sextet only.

The experiments on single crystal \cite{Oikawa} indicate that UGe2 has a base-centered 
orthorhombic crystal structure.
The sextet splits into doublet and quartet and the doublet's energy level is lower. 
Then one obtains two well separated systems of $f$ electrons, which are two sources of magnetism in $UGe_2$. This
 justifies the consideration of an effective model in terms of two vector fields ${\bf M}_{1i}$ and
 ${\bf M}_{2i}$ which identify the local orientation of the magnetization of the different groups of $f$ electrons.
 \bea\label{effH}
 h_{eff} & = & -  J_1\sum\limits_{  \langle  ij  \rangle  } {{\bf M}_{1i} \cdot
{\bf M}_{1j}}-  J_2\sum\limits_{  \langle  ij  \rangle  } {{\bf M}_{2i}
\cdot {\bf M}_{2j}} \nonumber \\
 & - & J\sum\limits_i {{\bf
M}_{1i}}\cdot {\bf M}_{2i}.
\eea
The exchange constants $J_1,\,J_2$ and $J$ are positive (ferromagnetic), the
sums are over all sites of a three-dimensional cubic lattice, and
$\langle i,j\rangle$ denotes the sum over the nearest neighbors. The $LDA+U$ calculations show
the existence of well separated majority spin state with orbital projection $m_l=0$ \cite{Pickett1}.
This can be modeled with spin $1/2$ fermion and ${\bf M}_{2i}$ is the local magnetization of the itinerant
electron. The saturation magnetization $"m"$ is close to $1/2$ at ambient pressure and decreases with increasing the pressure.
This accounts for the fact that some sites, in the ground state, are doubly occupied or empty.
The contribution of the others
uranium $5f$ electrons, which occupy the lowest energy level bands, to the magnetization is described by
${\bf M}_{1i}$ vector with saturation magnetization $s=1$.
One thinks of these electrons as localized, but they are not perfectly localized in $UGe_2$. This means that
saturation magnetization $s$ could be smaller then one.

Renormalized spin-wave theory, which accounts for the magnon-magnon interaction,
and its extension are developed in the present paper, to describe two ferrimagnetic phases in
the system Eq.(\ref{effH}): low temperature phase  $0<T<T^{*}$ , where $<{\bf M}_{1i}>$ and
 $<{\bf M}_{2i}>$ contribute the ordered ferromagnetic
moment, and high temperature phase $T^{*}<T<T_C$, where only $<{\bf M}_{1i}>$ is nonzero.
Both phases are strictly ferromagnetic.

To proceed we use the Holstein-Primakoff representation of the spin vectors ${\bf
M}_{1j}(a^+_j,a_j)$ and ${\bf M}_{2j}(b^+_j,\,b_j)$, where $a^+_j,\,a_j$
and $b^+_j,\,b_j$ are Bose fields. One represents the Hamiltonian Eq.(\ref{effH}) in terms of these
Bose fields keeping only the quadratic and quartic
terms. The next step is to represent the Hamiltonian in the Hartree-Fock  approximation
$h\approx h_{HF}=h_{cl}+h_q$, where
\bea\label{HF}
 h_{cl}& = & 3 N J_1 s^2 (u_1-1)^2+ 3 N J_2 m^2 (u_2-1)^2 \nonumber \\
& + &  N J s m (u-1)^2, \\
 h_{q} & = & \sum\limits_{k}\left (\varepsilon^a_k\,a_k^+a_k\,+\,\varepsilon^b_k\,b_k^+b_k\,-
 \,\gamma\,(a_k^+b_k+b_k^+a_k)\,\right).\nonumber \eea
 In equation (\ref{HF}) the wave vector $k$ runs over the first
Brillouin zone of a cubic lattice, $N$ is the number of lattice's sites,
$u_1,\,u_2,\,u$ are the Hartree-Fock parameters, $\gamma = J\,u\,\sqrt{s\,m}$ and
 the dispersions are given by the equalities
 \be\label{1dispersion}
\varepsilon^a_k =  2s\,J_1\,u_1\varepsilon_k
\,+\,m\,J u \qquad
\varepsilon^b_k = 2m \,J_2\,u_2 \varepsilon_k \,+\,s\,J\,u \ee
with $\varepsilon_k=3-\cos k_x-\cos k_y-\cos k_z$. Equation (\ref{1dispersion}) shows that
the Hartree-Fock parameters $u_1,\,u_2$ and $u$ renormalize
the exchange constants $J_1,\,J_2$ and $J$ respectively.

To diagonalize the Hamiltonian, one introduces new Bose fields
$\alpha_k,\,\alpha_k^+,\,\beta_k,\,\beta_k^+$,
\be\label{Bogoltrans}
a_k  =  \cos\theta_k\,\alpha_k\,+\,\sin\theta_k\,\beta_k,  \qquad
b_k  = -\sin\theta_k\,\alpha_k\,+\,\cos\theta_k\,\beta_k
\ee
with coefficients of transformation, \be\label{Bogotrans2}
\cos\theta_k = \sqrt{\frac 12\,\left (1+\frac
{\varepsilon^a_k-\varepsilon^b_k}{\sqrt{(\varepsilon^a_k-\varepsilon^b_k)^2+4\gamma^2}}\right
)},\ee
and $\sin\theta_k  =  \sqrt{\left (1-\cos^2\theta_k\right)}$.
The transformed
Hamiltonian adopts the form \be \label{letter23} h_{q} = \sum\limits_{k}\left
(E^{\alpha}_k\,\alpha_k^+\alpha_k\,+\,E^{\beta}_k\,\beta_k^+\beta_k\right),
\ee with new dispersions \bea  \label{letter24}
E^{\alpha}_k  & = & \frac 12\,\left
[\varepsilon^a_k+\varepsilon^b_k\,+ \,
\sqrt{(\varepsilon^a_k-\varepsilon^b_k)^2+4\gamma^2}\right] \nonumber \\
\\
E^{\beta}_k  & = & \frac 12\,\left
[\varepsilon^a_k+\varepsilon^b_k\,- \,
\sqrt{(\varepsilon^a_k-\varepsilon^b_k)^2+4\gamma^2}\right] \nonumber \eea

With positive exchange constants $J_1,\,J_2,J$ and positive Hartree-Fock parameters
$u_1,\,u_2,\,u$ the Bose fields'
dispersions are positive $\varepsilon^a_k>0,\,\,\varepsilon^b_k>0$
for all values of $k\in B$. As a result, $E^{\alpha}_k>0$ and
$E^{\beta}_k\geq 0$ with $E^{\beta}_0=0$. Near the zero wave vector,
$E^{\beta}_k\approx \rho k^2$ where $\rho$ is the spin-stiffness constant.
Hence, $\beta_k$ is the long-range \textbf{(magnon)} excitation in the effective
theory, while $\alpha_k$ is a gapped excitation.

The system of equations for the Hartree-Fock parameters
have the form
\bea\label{HFeq} u_1 & = & 1-\frac {1}{3s} \frac 1N \sum\limits_{k} \varepsilon_k \left[\cos^2\theta_k \,n_k^{\alpha}\, +\, \sin^2 \theta_k\, n_k^{\beta}\right]\nonumber \\
u_2 & = & 1-\frac {1}{3 m} \frac 1N \sum\limits_{k} \varepsilon_k \left[\sin^2\theta_k \,n_k^{\alpha}\, +\, \cos^2\theta_k\, n_k^{\beta}\right]\nonumber \\
u & = & 1-\frac 1N \sum\limits_{k}\left[\left(\frac {1}{2s}\cos^2\theta_k+\frac {1}{2m}\sin^2\theta_k \right)\,n_k^{\alpha}\right.\nonumber \\
& + & \left.\left(\frac {1}{2m}\cos^2\theta_k+\frac {1}{2s}\sin^2\theta_k \right) \,n_k^{\beta} \right. \\
& + & \left. \frac {J u}{\sqrt{(\varepsilon^a_k\,-\,\varepsilon^b_k)^2\,+\,4\gamma^2}} \,(n_k^{\alpha}-n_k^{\beta})\right]\nonumber
\eea
where $n_k^{\alpha}$ and $n_k^{\beta}$ are the Bose functions of $\alpha_k$ and $\beta_k$ excitations.
The Hartree-Fock parameters, the solution of the system of equations (\ref{HFeq}), are positive functions of
$T/J$, $u_1(T/J)>0,\,u_2(T/J)>0$ and $u(T/J)>0$. Utilizing these functions, one can calculate
the spontaneous magnetization of the system, which is a sum of the spontaneous magnetization $M_1=<M^z_{1i}>$ and $M_2=<M^z_{i2}>$:
$M\,=\,M_1\,+\,M_2$. In terms of the Bose functions of the $\alpha_k$ and $\beta_k$ excitations they adopt the form
\bea\label{letter26}& & M_1\,=\,s-\frac 1N \sum\limits_{k}\left
[\cos^2\theta_k\,n^{\alpha}_k\,+\,\sin^2\theta_k\,n^{\beta}_k
\right], \nonumber
\\
& & M_2\,=\,m-\frac 1N \sum\limits_{k}\left
[\sin^2\theta_k\,n^{\alpha}_k\,+\,\cos^2\theta_k\,n^{\beta}_k\right],\nonumber \\
& & M\,=\,s\,+\,m-\frac 1N \sum\limits_{k}\left
[n^{\alpha}_k\,+\,n^{\beta}_k\right].\eea

Calculating the spontaneous magnetization one obtains that at characteristic temperature $T^*$
the spontaneous magnetization $M_2$ becomes equal to zero, while the spontaneous magnetization
$M_1$ is still nonzero.
This is because the magnon excitation $\beta_k$ in the effective theory
Eq.(\ref{effH}) is a complicated mixture of the transversal
fluctuations of the vectors ${\bf M}_{i1}$ and ${\bf M}_{i2}$
Eq.(\ref{Bogoltrans}). As a result, the magnons' fluctuations suppress
in a different way the different magnetic orders.
Above $T^*$ the system of equations (\ref{HFeq}) has no solution and one has to modify
the renormalized spin-wave theory.

To formulate mathematically the modified RSW theory one introduces \cite{Karchev08a}
two parameters $\lambda_1$ and $\lambda_2$ to enforce the two magnetic
moments to be equal to zero in paramagnetic phase. To this end, we add two new terms
to the effective Hamiltonian Eq.(\ref{effH}), \be
\label{letter28} \hat{h}_{eff}\,=\,h_{eff}\,-\,\sum\limits_i \left
[\lambda_1 M^z_{1i}\,+\,\lambda_2 M^z_{2i}\right]. \ee
In Hartree-Fock approximation, in momentum space, the Hamiltonian adopts the form
\be \label{letter18}
 \hat {h}_{q} = \sum\limits_{k}\left (\hat{\varepsilon}^a_k\,a_k^+a_k\,+\,\hat{\varepsilon}^b_k\,b_k^+b_k\,-
 \,\gamma\,(a_k^+b_k+b_k^+a_k)\,\right),\ee
 where the the new dispersions are
 \be \label{dispersion2}
\hat{\varepsilon}^a_k\,=\varepsilon^a_k\,+\,\lambda^l, \qquad
\hat{\varepsilon}^b_k\,=\varepsilon^b_k\,+\,\lambda^{it}.\ee

Utilizing the same transformation Eq.(\ref{Bogoltrans}) with coefficients
$\cos\hat{\theta}_k$ and $\sin\hat{\theta}_k$ expressed by means of $\hat{\varepsilon}^a_k$ and
$\hat{\varepsilon}^b_k$
one obtains the Hamiltonian in diagonal form (\ref{letter23}) with dispersions\,\,
$\hat{E}^{\alpha}_k =\hat{E}^{+}_k$ and $\hat{E}^{\beta}_k =\hat{E}^{-}_k$,
where
$\hat{E}^{\pm}_k  = \frac 12\,\left
[\hat{\varepsilon}^a_k+\hat{\varepsilon}^b_k\,\pm \,
\sqrt{(\hat{\varepsilon}^a_k-\hat{\varepsilon}^b_k)^2+4\gamma^2}\right]$.

It is convenient to represent the $\lambda$
parameters in the form
$\lambda_1 = m J (\mu_1 - 1), \quad \lambda_2 = s J
(\mu_2 - 1)$.
In terms of the $\mu$ parameters the dispersions adopt the form
$\hat{\varepsilon}^a_k  = 2sJ_1\,u_1\varepsilon_k+mJ u\mu_1, \quad
\hat{\varepsilon}^b_k = 2mJ_2u_2\varepsilon_k+sJu\mu_2$
The renormalized spin-wave theory is reproduced when
$\mu_1=\mu_2=1$($\lambda_1=\lambda_2=0$). We assume $\mu_1$
and $\mu_2$ to be positive. Then,
$\hat{\varepsilon}^a_k>0$, $\hat{\varepsilon}^b_k>0$, and
$\hat{E}^{\alpha}_k>0$ for all values of the wave-vector $k$.
The $\beta_k$ dispersion is
non-negative, $\hat{E}^{\beta}_k\geq0$ if $\mu_1 \mu_2\geq1$. In
the particular case $\mu_1 \mu_2=1$\,\, $\hat{E}^{\beta}_0=0$
and near the zero wave vector $\hat{E}^{\beta}_k\approx \hat{\rho}
k^2$, so $\beta_k$ boson is the long-range excitation (magnon) in the
system. In the case $\mu_1 \mu_2>1$, both $\alpha_k$ boson and
$\beta_k$ boson are gapped excitations.

The parameters $\lambda_1$ and $\lambda_2$ ($\mu_1, \mu_2$) are introduced
to enforce the spontaneous magnetizations $M_1$ and $M_2$
to be equal to zero in the paramagnetic phase. One finds out the parameters
$\mu_1$ and $\mu_2$, as well as the Hartree-Fock parameters, as functions of temperature,
solving the system of five equations: equations (\ref{HFeq}) and the equations $M_1=M_2=0$,
where the  spontaneous magnetizations have the same representation as equations (\ref{letter26})
but with coefficients $\cos\hat{\theta}_k,\,\,
\sin\hat{\theta}_k$, and dispersions $\hat{E}^{\alpha}_k,\,\,
\hat{E}^{\beta}_k$ in the expressions for the Bose functions.
The numerical calculations show that for high enough temperature
$\mu_1\mu_2>1$. When the temperature decreases the product $\mu_1\mu_2$ decreases,
remaining larger than one. The temperature at which the product becomes equal
to one ($\mu_1\mu_2=1$) is the Curie temperature.

Below $T_C$, the spectrum contains magnon
excitations, thereupon $\mu_1\mu_2=1$. It is convenient to
represent the parameters in the following way:\,\,
$\mu_2=\mu, \quad\mu_1=1/\mu$.

In the ordered phase magnon excitations are the origin of the suppression of the magnetization.
Near the zero temperature their contribution is small and at zero temperature
spontaneous magnetizations $M_1$ and $M_2$ reach their saturations $(M_1=s, \,\,M_2=m)$.
On increasing the temperature magnon fluctuations suppress the different ordered moments
in different way.
At $T^*$ the spontaneous magnetization $M_2$ becomes equal to zero.
Increasing the temperature above $T^*$, $M_2$ should be zero. This is why we impose the condition
$M_2(T)=0$ if $T>T^{*}$. For temperatures above $T^*$, the parameter $\mu$ and
the Hartree-Fock parameters are solution of a system of four equations, equations (\ref{HFeq})
with $\cos\hat{\theta}_k,\,\,\sin\hat{\theta}_k,\,\,\hat{\varepsilon}_k^a,\,\,\hat{\varepsilon}_k^b,\,\,\hat{E}^{\alpha}_k,\,\,\hat{E}^{\beta}_k$ instead of $\cos\theta_k,\,\,\sin\theta_k,\,\,\varepsilon_k^a,\,\,\varepsilon_k^b,\,\,E^{\alpha}_k,\,\,E^{\beta}_k$,
and the equation $M_2=0$.
One utilizes the
obtained functions $\mu(T)$, $u_1(T)$, $u_2(T)$, $u(T)$ to calculate the spontaneous magnetization
as a function of the temperature. Above $T^*$, the magnetization of
the system is equal to $M_1$.

The resultant magnetization-temperature
curves, for different choices of the model parameters, are depicted in figure. I set the Curie temperature
to be equal to the experimental one. This fixes the exchange constant $J$. The constants $J_1/J$ and $J_2/J$
are chosen so that the ratio $T_C/T^*$ to be close to the experimental value.
\begin{center}
\begin{figure}[htb]
\label{letterfig5}
\centerline{\psfig{file=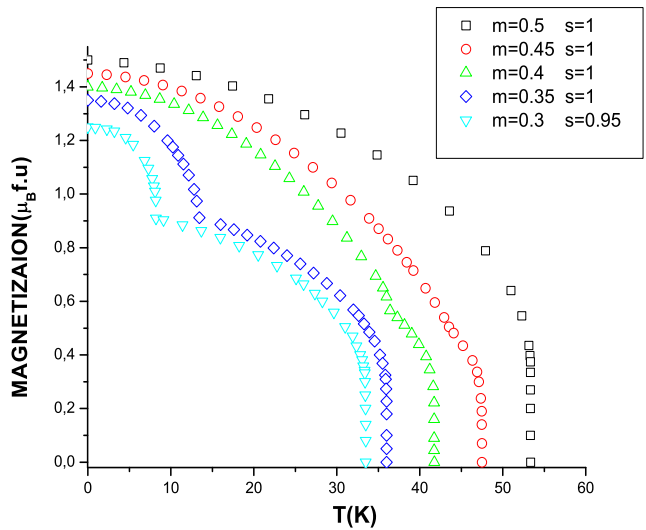,width=9.5cm,height=8cm}}
\caption{(color online)\,\,Magnetization-temperature curve obtained within an effective two
magnetic ordered moments model of $UGe_2$ magnetism}
\end{figure}
\end{center}
The first curve from above (black squares) is calculated for parameters $m=0.5,\,s=1,\,J_1/J=0.05$ and $J_2/J=0.0005$.
The strong interaction between itinerant and "localized" electrons aligns their magnetic orders
so strong  that they become zero at one and just the same temperature $T_C$. The magnetization-temperature curve
is typical Curie-Weiss curve. The result is different if the exchange constant $J$ is relatively smaller.
The ferromagnetic phase is divided into two phases: low temperature phase $0<T<T^*$ where $M_1$ and $M_2$
give contribution to the magnetization, and high temperature ferromagnetic phase $T^*<T<T_C$
where $M_2=0$.
The next curve (red circles) is obtained for parameters $m=0.45,\,s=1,\,J_1/J=0.16,\,J_2/J=0.0016$,
the third one (green triangles) for parameters  $m=0.4,\,s=1,\,J_1/J=0.18$ and $J_2/J=0.0018$,
the fourth curve (blue rhombs) corresponds to parameters  $m=0.35,\,s=1,\,J_1/J=0.4$ and $J_2/J=0.004$,
and for the last one $m=0.3,\,s=0.95,\,J_1/J=0.57$ and $J_2/J=0.0057$.
The curves show that increasing the constants $J_1/J$ and $J_2/J$ the ration $T_C/T^*$ increases ($T_C/T^*=1,\,1.092,\,1.46,\,2.68
,\,4.08$), and
$T^*$ approaches to zero ($T^*=53.35K,\,43.511K,\,36.433K,\,13.44K,\,8.21K$).
Comparing with experiment \cite{2fmp3,2fmp6} one concludes that increasing the pressure
the exchange constant $J$ increases, but exchange constants $J_1$ and $J_2$ increase faster,
so that the ratios  $J_1/J$ and $J_2/J$ increase.

The anomalous temperature dependence
of the ordered moment, known from the experiments with $UGe_2$
\cite{2fmp2,2fmp5,2fmp6,2fmp7}, is very well reproduced theoretically.
Below $T_x$ ($T^*$ in the present paper) the ferromagnetic
moment increases in an anomalous way. The low temperature, large moment phase is referred to as $FM2$,
while the high temperature low-moment phase is referred to as $FM1$ \cite{2fmp7,Pfleiderer2}. Both phases are strictly
ferromagnetic in accordance with experiment \cite{Huxley03b}.
The present theoretical result gives new insight into $FM1\rightarrow FM2$ transition.
It is shown that
between Curie temperature and $T^*<T_C$ the contribution of the itinerant $UGe_2$ electrons to the magnetization is zero.
They start to form magnetic moment at $T^*$.

There are experiments which support the present theoretical result. The measurements \cite{2fmp5}
show that the resistivity display a down-turn around $T_C$ and  $T^*(=T_x)$.
The last one is best seen in terms of a broad
maximum in the derivative $d\rho/dT$ \cite{Oomi1}. It is well known that the onset of magnetism in the itinerant
systems is accompanied with strong anomaly in resistivity \cite{2fmp12}. The experiments  \cite{2fmp5,Oomi1} prove that
there are two groups of $5f$ uranium electrons. One of them starts to form magnetic order at Curie temperature,
the other one does this at temperature $T_x(=T^*)$ well below $T_C$, in agreement with the theoretical result.
Further evidence for the nature of the
$FM1\rightarrow FM2$ transition has been observed in the high resolution photoemission, which show
the presence of a narrow peak in the density of states below $E_F$ that suggests itinerant
ferromagnetism \cite{Ito}.

In summary, it is shown that the anomalous temperature dependence
of the ordered moment is a result of the splitting of  $5f$ uranium electrons into two groups due to the strong
spin-orbit coupling. c

It is impossible to require
the theoretically calculated Curie temperature and magnetization-temperature curves to be in
exact accordance with experimental results. The models are idealized, and they do not consider
many important effects. Because of this it is important to formulate theoretical criteria for
adequacy of the method of calculation.
In my opinion the calculations should be in accordance with the Mermin-Wagner theorem \cite{M-W}.
It claims that at nonzero temperature, a one-dimensional or two-dimensional isotropic
spin-S Heisenberg model with finite-range exchange interaction can be neither ferromagnetic
nor antiferromagnetic. The renormalized spin-wave theory, developed in the present paper,
being approximate captures the essentials of the magnon
fluctuations and satisfies the Mermin-Wagner theorem.

This work was partly supported by a Grant-in-Aid DO02-264/18.12.08 from NSF-Bulgaria.

\end{document}